\documentclass[10pt, conference,compsocconf]{IEEEtran}  

\usepackage[bottom]{footmisc}
\usepackage{ifpdf}
\ifCLASSINFOpdf        
   \usepackage[pdftex]{graphicx}
\else
   \usepackage[dvips]{graphicx}
\fi
\usepackage[cmex10]{amsmath}   
\usepackage{algorithmic}
\usepackage{array}
\usepackage{bm}
\usepackage{mdwmath}
\usepackage{mdwtab}
\usepackage{eqparbox}
\usepackage{url}
\usepackage{subfigure}
\usepackage{balance}
\usepackage{multirow}
\usepackage{array}       
\usepackage[numbers,square,comma,sort&compress]{natbib}   
\usepackage[lined, algonl, boxed]{algorithm2e}
\usepackage{booktabs}
\usepackage{float}
\usepackage{dashrule}

\newcommand{\paragraphX}[1]{\vskip 4pt \noindent \textbf{#1} \hskip .05in}

\newcommand{\putk}{\bm{{\mathsf{put}}}}
\newcommand{\getk}{\bm{\mathsf{get}}}

\newcommand{\lookup}{\bm{\mathsf{lookup}}}
\newcommand{\whanau}{Wh\={a}nau }

\newcommand{\wiki}{\bm{\mathsf{wiki}}}

\newcommand{\flic}{\bm{\mathsf{flic}}}
\newcommand{\cat}{\bm{\mathsf{cat}}}

\newcommand{\ham}{\bm{\mathsf{ham}}}
\newcommand{\fb}{\bm{\mathsf{fb}}}
\newcommand{\astro}{\bm{\mathsf{astro}}}

\usepackage[final,inline,nomargin]{fixme}      
\usepackage{color}
\usepackage{soul}          

\newenvironment{smitemize}
  {\begin{list}{$\bullet$}
     {\setlength{\parsep}{0pt}
      \setlength{\leftmargin}{10pt}
      \setlength{\topsep}{-7pt}
      \setlength{\labelwidth}{5pt}
      \setlength{\itemsep}{1pt}}}
  {\end{list}}

\begin{document}

\title{iPersea : The Improved Persea with Sybil Detection Mechanism}

\author{
      \IEEEauthorblockN{Mahdi Nasrullah Al-Ameen and Matthew Wright\\}
      \IEEEauthorblockA{Department of Computer Science and Engineering\\
                 The University of Texas at Arlington \\
                 Arlington, TX, USA\\
              mahdi.al-ameen@mavs.uta.edu, mwright@cse.uta.edu}}

\maketitle

\begin{abstract}
P2P systems are highly susceptible to Sybil attacks, in which an attacker creates a large number of identities and uses them to control a substantial fraction of the system. Persea is the most recent approach towards designing a social network based Sybil-resistant DHT. Unlike prior Sybil-resistant P2P systems based on social networks, Persea does not rely on two key assumptions: (i) that the social network is fast mixing, and (ii) that there is a small ratio of {\em attack edges} to honest peers. Both assumptions have been shown to be unreliable in real social networks. The hierarchical distribution of node IDs in Persea confines a large attacker botnet to a considerably smaller region of the ID space than in a normal P2P system and its replication mechanism lets a peer to retrieve the desired results even if a given region is occupied by attackers. However, Persea system suffers from certain limitations, since it cannot handle the scenario, where the malicious target returns an incorrect result instead of just ignoring the lookup request. In this paper, we address this major limitation of Persea through a Sybil detection mechanism built on top of Persea system, which accommodates inspection lookup, a specially designed lookup scheme to detect the Sybil nodes based on their responses to the lookup query. We design a scheme to filter those detected Sybils to ensure the participation of honest nodes on the lookup path during regular DHT lookup. Since the malicious nodes are opt-out from the lookup path in our system, they cannot return any incorrect result during regular lookup. We evaluate our system in simulations with social network datasets and the results show that catster, the largest network in our simulation with $149700$ nodes and $5449275$ edges, gains $100$\% lookup success rate, even when the number of attack edges is equal to the number of benign peers in the network.

\end{abstract}


\section {Introduction}

Peer-to-peer (P2P) systems are inherently vulnerable to Sybil attacks, in which an attacker creates a large number of pseudonymous entities and use them to gain a disproportionately large influence over the system~\cite{sybil_kad,sybil_attack}. The attackers then collude to launch further attacks, such as taking over resources and disrupting connectivity to subvert the system's operation. Such attacks have been shown to be quite problematic in {\em structured} P2P systems in which nodes are placed into a distributed hash table (DHT) like Chord~\cite{chord}, CAN~\cite{CAN}, Pastry~\cite{pastry}, and Kademlia~\cite{kademlia}. Kademlia was the basis for both the Kad network and Vuze, DHTs used in the popular BitTorrent file-sharing P2P system with millions of users each. Researchers have documented this vulnerability in real-world systems, including the Maze P2P file-sharing system~\cite{lian_maze,yang_maze} and the Vanish data storage system~\cite{wolchoky_vanish}.

Recent research has focused on leveraging information from social networks to make the system robust against Sybil attackers, resulting in a number of decentralized apporaches~\cite{whanau1,whanau2,sybillimit,sybilguard,gatekeeper, sybildefender,xvine}. The key to these approaches is the idea that honest and malicious nodes can be effectively partitioned into two subgraphs in the social network. The link between an honest node and a malicious peer is called an attack edge, which represents an act of social engineering to convince the honest node to add the link.

These mechanisms are based on two key assumptions: (i) that the online social networks are {\em fast-mixing}, meaning that a random walk in the honest part of the network approaches the unifrom distribution in a small number of steps, and (ii) that the number of attack edges are rather limited in online social networks. Recent studies~\cite{mixing,viswanath_analysis,sybil_wild,id_theft}, however, show that the above assumptions do not hold in real-world social networks. So, it remain an open research problem, until Persea~\cite{persea} was proposed, to design a social network based Sybil defense mechanism, which does not rely on these assumptions. 

\subsection{Motivation}
Persea~\cite{persea} derives its Sybil resistance by assigning IDs through a bootstrap tree, the graph of how nodes have joined the system through invitations. Persea argues that building a bootstrap tree is more realistic than assuming that the clients have access to lists of social network connections from a system like Facebook. Also, IDs are certified in Persea, making attacks based on ID forging impossible outside of attacker-controlled ID ranges. In Persea, the (key,value) pair is replicated in evenly spaced nodes, so that even if a given region is occupied by the attacker, the desired (key,value) pair can be retrieved from other regions. We give a overview of Persea system in~\S\ref{persea}. 

The Persea approach offers a number of important advantages over previous schemes:

\begin{smitemize}
\item Unlike prior Sybil-resistant P2P systems based on social networks, Persea does not rely on two key assumptions: (i) that the social network is fast mixing, and (ii) that there is a small ratio of attack edges to honest nodes. 
\item The hierarchical distribution of node IDs limits the attackers to isolated regions in ID space.
\end{smitemize}
\vspace{0.2cm}

However, Persea system suffers from certain limitations that make its use questionable in real world scenario. In Persea DHT, a (key,value) pair is replicated in a number of nodes and the lookup operation is performed for each target node to get the value, associated with the search-key. Persea assumes that when the target node is malicious it does not return incorrect result, rather ignores the lookup request. So, In Persea DHT, if the initiator of a lookup retrieves the correct result from at least one benign target, the lookup is termed to be successful. But in real-world scenario, the malicious target may reply with incorrect result to make it harder for the initiator to retrieve the correct one from the set of different returned results. Persea system cannot handle such obvious attacks.

The simplest solution to this limitation of Persea could be implementing {\em majority voting} scheme, where the initiator picks the result with higher count. If the counts of two types of results are same, the initiator randomly picks one as the final result. In our simulation, we implement the above strategy of adversaries and evaluate Persea with majority voting scheme. However, our results show that the lookup success rate in Persea sharply decreases with the increase in attack edges (See~\S\ref{results}), which infers that majority voting is not effective enough to make the system robust, when the malicious target returns incorrect result. So, the efficacy of Persea is left as an open question in real-world scenario. 

\subsection{Contributions}
In this paper, we address this major limitation of Persea and develop a Sybil detection scheme on top of Persea system. We propose {\em inspection lookup} to detect the Sybils, which is a specially designed lookup mechanism to determine the {\em status} (honest or malicious) of a node. This lookup seems as a regular DHT lookup to a peer, whose status is being inspected, so that an attacker cannot play a fabricated role during inspection lookup to prove it as an honest node. We introduce the idea of {\em collaborative friends}, the groups of benign peers, who agree to execute the inspection lookups for detecting the Sybils. We provide a detailed description of our Sybil detection mechanism in~\S\ref{detection}.

We develop a mechanism to filter the detected Sybil nodes from the lookup path for ensuring the participation of benign peers during regular DHT lookup. We then incorporate our filtering scheme with the lookup mechanism in Persea. So, in our system the attackers are opt-out of the lookup path and thus, they can neither intercept a lookup (as an intermediate node) nor return an incorrect result (as the target node).

While we incorporate our Sybil detection mechanism with Persea, we name the new system {\em iPersea} (\textbf{i}mproved \textbf{Persea}), which inherits the advantages of Persea that it gains over prior social network based Sybil-resistant systems. So, iPersea does not depend on the fast-mixing social network and small ratio of attack edges to honest peers. We justify our claims through evaluations in~\S\ref{results}. We simulate for networks with different {\em clustering coefficients} to show that our system does not depend on the fast-mixing nature of a network. The clustering coefficient is a measure of degree to which nodes in a network tend to cluster together~\cite{measure,pattern}, and it is therefore directly related to the fast-mixing property of a network~\cite{pattern}. To validate the claim that iPersea does not depend on the small ratio of attack edges to honest peers, we evaluate for this ratio, upto $1.5$ and find consistent results, where the lookup success rate for any network in our evaluations is no less than $92.9$\%.

Our experimental results show that $70$\% of lookups succeed in Persea, while the ratio of attack edges to honest nodes is $0.5$ in facebook social network. However, in iPersea, $93.6$\% lookups succeed in facebook, even when the ratio of attack edges to honest nodes is increased to one. In flickr and catster (largest network in our simulation with $149700$ nodes), $100$\% lookups get successful in our system when we have equal number of attack edges and honest nodes. 

The remainder of the paper is organized as follows: we give a overview of Persea system in~\S\ref{persea} and explain the design our Sybil detection mechanism in~\S\ref{detection}. We then describe the attack model in~\S\ref{attack} before presenting our simulation results in~\S\ref{results}. In~\S\ref{related} we discuss the related works in Sybil defense. We give a direction to our future work~\S\ref{future} and then conclude in~\S\ref{conclusion}.

\section{Overview of Persea}\label{persea}

In this section, we briefly describe the design of Persea~\cite{persea}. We begin with the overview of the system and then address the ID space allocation, ID certification and key replication. We also describe the routing table organization and lookup mechanism of Persea.

\subsection{Design Overview} 
Persea consists of two layers: a social network layer (the bootstrap graph) and the DHT layer. In Persea, an {\em edge} refers to a link between two nodes in the DHT layer. The bootstrap network and DHT are simultaneously built starting with a set of bootstrap nodes. The bootstrap nodes are the initiators of the system. They are connected to each other in both the bootstrap network and the DHT. Node IDs in the DHT are assigned to the bootstrap nodes such that they are evenly spaced over the circular ID space. Thus, the ID space of the DHT is divided into one region for each bootstrap node. 

A new peer must join the Persea system through an invitation from an existing node in the network. In general, it is expected that a node that is invited is socially known to the inviting peer. When a node is invited, it not only becomes a part of the bootstrap network but also gets a node ID in DHT layer. The new node gets a chunk of node IDs that it can use to invite more nodes for joining the network. ID assignments and chunk allocations are put into certificates signed by the parent nodes, and certificates are stored in the DHT itself to allow for reliable distributed checking of the chain of certificates all the way up to the bootstrap nodes. 

The DHT layer of Persea is based on Kademlia~\cite{kademlia}, a DHT that is widely adopted for the BitTorrent file-sharing P2P system. The main difference in Persea is that IDs are replicated evenly around the ID space for greater resiliency given the ID distribution scheme. 

\subsection{Hierarchical ID Space} 
We now briefly describe how node IDs are distributed in Persea. Each bootstrap node has a contiguous range of node IDs called a \textit{chunk}, which includes the bootstrap node's ID. A bootstrap node divides its chunk of node IDs into sub-chunks based on the \textit{chunk-factor}, a system parameter.

When a bootstrap node invites a peer to join the system, it assigns a node ID to the joining node from one of its sub-chunks and also assigns the new node control over the rest of the sub-chunk for further distribution. The newly joined node becomes the authority for distributing node IDs from the given sub-chunk. Thus, once the joining node becomes a part of the system, it can invite more nodes to join the system. Based on the invitation-relationship among peers, a {\em bootstrap tree} is formed in which an inviter node is the parent of its invited peers. If the number of bootstrap node is more than one, then it would have a forest of trees, where each bootstrap node is the root of each tree. The chunk-factor and size of the ID space define the maximum possible height and width of a tree. This mechanism has the advantage that even if a bot compromises a node and leverages it to add a large number of malicious nodes to the system, they will be still confined in a particular region of ID space.

\subsection{Routing table organization} 
In the DHT layer, each node maintains a routing table of $b$ node lists for a $b$-bit ID space. Each list has up to $k$ entries and is called a {\em $k$-bucket}. Each $k$-bucket entry contains the IP address, port, node ID, and public key of another node. The list is organized so that the ID of a node in the $b^{th}$ list of a node with ID $i$ should share the first $b-1$ bits of $i$ and have a different $b^{th}$ bit from $i$.

\subsection{Replication} 
In Persea, a (key,value) pair is stored in evenly spaced nodes, so that even if a given region is occupied by the attackers, the desired (key, value) pair can be retrieved from other regions. In Persea, the ID space is virtually divided into $R$ regions and the (key,value) pair is replicated in $R$ evenly spaced nodes, one in each virtual region. The evaluations in Persea show that $R=7$ gives the optimal results for the networks, considered in their simulation~\cite{persea}.

\subsection{Lookup mechanism} 
Node lookup in Persea is initiated by the $\lookup(key)$ request where a node queries the $\alpha$ nodes in its $k$-buckets that are the closest ones to the desired key. Each of the $\alpha$ nodes sends the initiator $\beta$ node IDs from its $k$-bucket closest to the target node. From the set of returned node IDs, the initiator selects $\alpha$ nodes for the next iteration. This process is iterated until the target is found or no nodes are returned that are closer than the previous best results.

The initiator of a lookup performs $R$ such independent parallel lookup operations and when an owner is found from any of $R$ independent lookups, the initiator sends the owner a message for either the store ($\putk(key, value)$) or retrieval ($\getk(key)$) operation. We incorporate our Sybil-filtering scheme (see~\S\ref{mlook}) with this lookup mechanism to opt-out the attackers from the lookup path.

\section{Sybil Detection Mechanism}\label{detection}
Our Sybil detection mechanism detects an attacker by exploiting its malicious behavior during a lookup. We propose inspection lookup to detect the Sybils, which determines the status (honest or malicious) of a node based on its response to the lookup request. Inspection lookup accommodates certain strategies to appear as a regular lookup to a peer, whose status is being inspected, so that an attacker cannot fabricate its behavior during inspection lookup to prove it as a benign node. The status (honest or malicious) of a node, determined through inspection lookup, is used to filter Sybil nodes during regular lookup. We incorporate our filtering scheme with the lookup mechanism of Persea to ensure higher lookup success rate by ensuring the participation of honest peers on the lookup path. 

Our Sybil detection mechanism is based on the following assumptions:

\begin{smitemize}

\item The adversary intercepts as many lookup queries as possible. As a target node, an attacker does not return the correct value, associated with the search-key.
\item An honest node adheres to the protocol and acts legitimately.
\end{smitemize}
\vspace{0.2cm}

In our mechanism, the status of a node represents whether it is honest or malicious, represented by '+' (honest) or '-' (malicious). Each parent node determines the status of its direct children with the help of selected peers, called collaborative friends. 

\subsection{Selecting collaborative friends} A peer (say it node P) requests its parents, grandparents and other ancestor nodes to suggest trusted peers, who agree to be the collaborative friends of node P for detecting the Sybils. An ancestor can suggest any number of collaborative friends, depending upon number of peers it trust and their willingness to collaborate. The more collaborative friends, node P has from different layers of hierarchical ID space, the harder it is for the child (may be, an attacker) of node P to distinguish an inspection lookup from the regular one. Since, node P selects the initiator of an inspection lookup from the set of its collaborative friends, the randomness in the placement of collaborative friends in ID space contributes in rising the hurdles to distinguish between inspection and regular lookup. So, node P requests its ancestors at each upper layer to be its collaborative friends and suggest more trusted nodes. 

In ideal case, an ancestor node always returns trusted collaborative friends. However, in real-world scenario, an ancestor, which is not responsible enough in detecting the Sybils, may return randomly-selected collaborative friends. In our experiments, we consider both scenarios and the results for Sybil-detection and lookup success rate show very subtle differences between two approaches. So, random selection of collaborative friends can be effective in detecting the Sybils and consequently gaining high lookup success rate.

\subsection{Inspection lookup} The goal of inspection lookup is to detect the Sybils based on their responses to the lookup messages. To design an inspection lookup, we adapt the basic lookup mechanism of Persea (see~\S\ref{persea}) and incorporate following strategies so that an inspection lookup appears to be a regular lookup to the peer and consequently an attacker cannot distinguish it from a regular lookup. 

\begin{smitemize}
\item The source and target of an inspection lookup and also the node, whose status is to be inspected, are randomly selected.
\item Inspection lookups are performed at uniformly distributed random interval.
\item During the lookup operation in DHT, the role of a peer on the lookup path may be an intermediate hop or the target. In each inspection lookup, it is randomly selected which role (intermediate hop or target) of a peer would be inspected.
\end{smitemize}

\vspace{0.2cm}
While inspecting the role of a child as an intermediate hop, node P selects a peer (say it node F) from its list of collaborative friends to be the initiator of the inspection lookup. It selects one of its direct children (say it node T) as the target node. These selections are made randomly. From the set of direct children, whose status are not inspected yet, node P randomly selects a child (say it node C) as an intermediate node of the lookup. Based on the success of inspection lookup, the status of node C is determined.

According to the suggestion of node P, node F sends the lookup request to node C. Since the inspection lookup appears as a regular lookup to Node C, it follows the mechanism of regular lookup and returns $\beta$ nodes from its $k$-buckets that are closest ones to the search-key. If the set of returned nodes does not include node T, node F then sends lookup request to each of these $\beta$ nodes in the next iteration. This process is iterated until the target is found or no nodes are returned that are closer than the previous best results.

In the ID space of Persea, where the node IDs are hierarchically distributed, there is an obvious lookup path from node C to node T through their parent node P. So, if the inspection lookup does not reach node T, node C is considered responsible for this failure and gets '-' status. Node C gets '+' status when the lookup succeeds, since an honest node directs a lookup towards the target.

When the role of a peer as the target node is inspected, a randomly selected collaborative friend of node P (say it node $F_1$) stores a (key,value) pair in node C and makes sure that the key matches with the node ID of node C. After a random interval, the lookup request is sent to node C from a collaborative friend of node P (say it node $F_2$) to return the value associated with the search-key. Node $F_2$ is informed by node P about the desired value that should be returned by node C. If it is not returned, node C is marked with '-' status. If node C returns the correct value, it gets '+' status, since an honest peer always returns the appropriate value associated with the search-key.

\paragraphX{False positive and false negative.} During inspection lookup, an honest intermediate node, whose status is being inspected, may unintentionally return malicious peers that are closest to the target, increasing the probability of a lookup to fail. If the lookup fails, the honest node is marked with '-' status that increases the rate of false positive.

The rate of false positive is zero when the role of a peer as the target node is inspected, as an honest node always returns the correct value associated with the search-key. The rate of false negative is zero in both cases, since our mechanism correctly identifies an attacker exploiting its malicious response to the inspection lookup message.

The above discussions on false-positive and false-negative are applicable to the scenarios, which abide by the assumption of selecting trusted nodes only, as the collaborative friends. However, for random selection of collaborative friends an attacker may get selected. In this case, if the attacker initiates an inspection lookup as a collaborative friend, a peer, whose status is inspected, gets '-' status if it is honest and is marked with '+' status if it is malicious, irrespective of the outcome of inspection lookup. Thus, for random selection of collaborative friends, the rate of both false-positive and false-negative may get increased.

\subsection{Sybil-filtering Mechanism}\label{mlook} 
In this section, we describe the mechanism to opt-out the Sybils from the lookup path for ensuring the participation of honest nodes during regular lookup. In our evaluations, we incorporate our filtering mechanism with the basic lookup scheme of Persea.

During regular lookup, before selecting a peer (say it node Q) as an intermediate hop, the initiator of the lookup (say it node L) asks the parent of node Q to send its status. We assume, a malicious peer invites other attackers to join the network and promotes its children by giving '+' status. According to these assumptions, a malicious node gets '-' status only from a benign parent. So, if the status of node Q is found '+', node L gets the status of node Q's parent. This process is repeated for the other ancestors of node Q until the bootstrap node is reached or the '-' status is found for an ancestor. 

If the bootstrap layer is reached, it suggests that none of node Q's ancestors is marked with '-' status. So, node Q is considered as a benign peer and gets selected for the lookup. If '-' status is found for node Q or any of its ancestor nodes, node Q is termed as a malicious peer and does not get selected as an intermediate node for the lookup. Once node L gets the status of node Q, it stores that status to be used in future lookup.

Node L follows the same procedure, as described above, to get the status of a target node, for deciding whether to accept the value, returned by that node.

\section{Attack Model}\label{attack}

We inherit most of the features of the attack model in Persea~\cite{persea}, where attackers use social engineering to create attack edges in the social network. When a malicious peer joins the network, it gets a chunk of node IDs for further distribution and we assume, an attacker invites only malicious peers for joining the network to infiltrate the system with as many attackers as possible. Also, the attacker promotes its children by assigning '+' (honest) status without performing any inspection lookup for them. 

As in Persea, we assume that the attackers know the IDs of all other attackers and store only the information of malicious nodes in their $k$-buckets. The goal of the adversary is to intercept as many lookup queries as possible. During lookup, If an attacker gets the $\lookup$ message, it returns the node IDs of malicious peers from its $k$-bucket. When the attacker receives a $\getk$ message, instead of just ignoring the message (as in Persea~\cite{persea}), it returns an incorrect value in our attack model.

In our Sybil detection mechanism, for random selection of collaborative friends, an attacker (say it node A) may be selected as the collaborative friend. We assume, as the collaborative friend of a node (say it node P), when node A initiates an inspection lookup to inspect the status of a child of node P (say it node C), whatever be the results of lookup; if node C is a malicious peer, node A informs node P that the lookup succeeds, however, if node C is an honest peer, the response of node A to node P says, the lookup fails.
\begin{figure*}[!t]
\centering
\subfigure[Average hop-count per lookup]{
\label{fig:analysis_path}
\includegraphics[width=55mm,height=38mm]{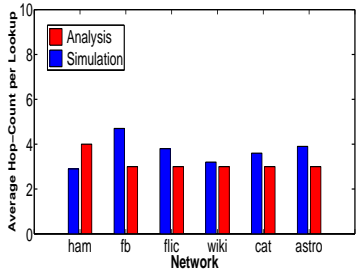}}\hspace{0.4cm}
\subfigure[Rate of false positive (trusted collaborative friends)]{
\label{fig:analysis_fpt}
\includegraphics[width=55mm,height=38mm]{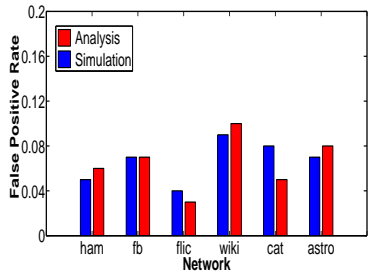}}
\subfigure[Rate of false positive (randomly selected collaborative friends)]{
\label{fig:analysis_fpr}
\includegraphics[width=55mm,height=38mm]{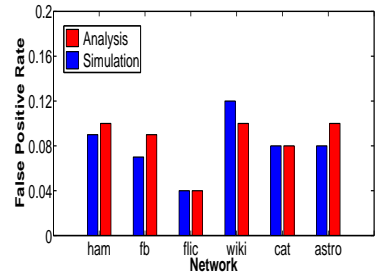}}\hspace{0.4cm}
\caption{Comparison between results from analysis and simulation}
\label{fig:analysis}
\end{figure*}

\section{Analysis}\label{analysis}

In this section, we develop analytical models to estimate average hop-count per regular lookup and the rate of false positive and false negative in our Sybil detection mechanism. 

\subsection{Lookup path length} We develop an analytical model to estimate the average hop-count per lookup. Let $e_p$ represent the average edges per node in a network and $a_h$ be the number of attack edges per honest node. So, $a_h=g/n$, where $g$ represents the number of attack edges and $n$ is number of honest nodes. When a lookup operation starts in our system, the initiator of the lookup selects $\alpha$ nodes from its routing table. So, the number of malicious peers in the set of $\alpha$ nodes is represented by $\alpha \times \frac{a_h}{e_p}$. In the next iteration, each of $\alpha$ nodes returns $\beta$ closest nodes (to the target) and a malicious peer always returns adversary nodes. Let $m_i$ represent the number of malicious nodes that the initiator gets in the set of $\alpha \times \beta$ nodes in $i^{th} (i\ge 1)$ iterations.  So, in the first iteration, the number of malicious peers that the initiator gets in the set of $\alpha \times \beta$ nodes is: $p_{m_1}=\left(\alpha \times \frac{a_h}{e_p}\times \beta \right) +\left((1-\alpha \times \frac{a_h}{e_p})\times \beta \times \frac{a_h}{e_p}\right)$. 

The attackers try to subvert the lookup query and thus the number of attackers in each iteration contributes to estimate the probability of a lookup to succeed. The number of malicious nodes, selected by the initiator in the set of $\alpha$ nodes for $(i+1)^{th}$ iteration is: $\alpha \times \frac{m_i}{\alpha \times \beta}=\frac{m_i}{\beta}$. So, the probability ($p_{m_i}$) of choosing a malicious nodes from the set of $\alpha$ nodes at iteration $i$ is represented by $\frac{m_i}{\alpha \times \beta}$. Thus, at any iteration $j (j\ge 1)$, we get $p_{m_j}$ by evaluating $\prod_{i=1}^{j} {p_{m_i}}$. The lookup continues until the target node is found and we estimate the probability of lookup failure at iteration $j$ by evaluating $p_{m_j}$. In our analysis, to estimate the path length of a lookup, we take a constant $l_c$ representing a very low probability of lookup failure and then we calculate the minimum value of $i$ as the lookup path length, where $p_{m_i}\le l_c$.

We consider $l_c = 0.001$ to get the lookup path length from our analytical model. To compare our analytical results with the results from our evaluations, we consider $a_h=1.0$, $\alpha = 5$ and $\beta = 7$ (as used in our simulations). The value of $e_p$ depends upon individual network. Figure~\ref{fig:analysis_path} shows the differences between our analysis and simulation for average hop-count per regular lookup. We find that our analytical estimations are very close to the experimental results. 

\subsection{Rate of false positive} To estimate the rate of false positive, we use the above analytical model for measuring $p_{m_j}$. We consider both trusted and randomly selected collaborative friends in our analysis. For trusted collaborative friends, the rate of false positive is zero when the role of a peer as the target is inspected. However, when the role of a peer as the intermediate node is inspected, we may get a rate of false positive, since an attacker may intercept the lookup (see ~\S\ref{detection} for explanation). So, the number of attackers, selected in an iteration is directly related to the rate of false positive. Thus, we use $p_{m_j}$ to estimate the false positive rate  and put $j=1$ in this case. In our analysis, we assume that $j=1$ is a reasonable estimation for hop-count per inspection lookup and also, we get the same value for $j$, when we take the {\em floor} of the results for average hop-count per inspection lookup in our evaluations (see Table~\ref{tab:hpc}).

While comparing our analytical estimations for false positive rate with the experimental results, we consider the ratio of attack edges to honest nodes is one. Figure~\ref{fig:analysis_fpt} illustrates the results for this comparison. We find that the difference between analysis and simulation is $0.01$ in the network datasets of hamsterster (\ham), flickr (\flic), wiki-Vote (\wiki) and ca-AstroPh (\astro). In the social network dataset of facebook (\fb), we get exactly same results from our analytical model and simulations. 

\begin{table*}[!t]
\renewcommand{\arraystretch}{1.3}
\caption{Topologies}
\centering
\begin{tabular}{|ll|l|c|c|c|}
\hline
\multicolumn{2}{|c|}{Network} & \multicolumn{1}{|c|}{Description} & Nodes & Edges & Avg. Clustering Coeff. \\
\hline
\hline
hamsterster & (\ham) & Social network of hamsterster.com &$2426$ & $16631$ & $0.08$\\
\hline
facebook & (\fb) & Social network of facebook.com &$63731$ & $1545686$ & $0.15$\\
\hline
flickr & (\flic)        & Network dataset of flickr.com & $80513$ & $5899882$ & $0.17$ \\
\hline 
wiki-Vote & (\wiki)     & Wikipedia who-votes-on-whom network & $7115$& $103689$ & $0.21$\\
\hline
catster & (\cat)        & Social network of Catster.com& $149700$ & $5449275$ & $0.43$\\
\hline
ca-AstroPh & (\astro)      & Collaboration network of Arxiv Astro Physics& $18772$ & $396160$ & $0.63$ \\
\hline
\hline
\end{tabular}
\label{tab:topology}
\end{table*}

Now, we estimate the false positive rate in a scenario, where the collaborative friends are randomly selected. In this case, the false positive rate may get increased not only by the attackers on the lookup path, but also by malicious collaborative friends. An attacker, as a collaborative friend, contributes to increase the false positive rate when it is selected to initiate an inspection lookup and the peer, whose status is being inspected in that lookup, is an honest one. So, to estimate the false positive rate, we measure the probability of selecting a malicious collaborative friend and an honest node (to inspect its status) in the same inspection lookup and then take the union of this probability with $p_{m_j} (j=1)$.

Let a node N be at level $l_h+1$ of the hierarchical ID space and $c_{nl}$ be the number of collaborative friends from each of its upper level. For simplicity of analysis, we consider $c_{nl}$ is same for each level. In our analysis, $e_p$ represents the average edges per node in a network and $a_h$ is the number of attack edges per honest node. So, when a node is randomly selected as a collaborative friend, $\frac{a_h}{e_p}$ represents its probability to be malicious. Thus, the expected number of malicious collaborative friends of node N is: $\frac{a_h}{e_p}\times l_h\times c_{nl}$. When node N randomly selects a peer from its set of collaborative friends to initiate an inspection lookup, $\frac{\frac{a_h}{e_p}\times l_h\times c_{nl}}{l_h \times c_{nl}}=\frac{a_h}{e_p}$ represents the probability of this collaborative friend to be an attacker. 
To inspect the status, the probability of selecting an honest child of node N is: $\frac{e_p-a_h}{e_p}$. Now, we measure the probability of selecting a malicious collaborative friend and an honest node (to inspect its status) in the same inspection lookup as: $\frac{a_h}{e_p}\times \frac{e_p-a_h}{e_p}$. So, for random selection of collaborative friends, we calculate $(\frac{a_h}{e_p}\times \frac{e_p-a_h}{e_p})+p_{m_j}-(\frac{a_h}{e_p}\times \frac{e_p-a_h}{e_p}\times p_{m_j})$ for the estimation of false positive rate. 

We compare our analytical estimations with the results from our simulation, shown in Figure~\ref{fig:analysis_fpr}, where $g/n=1.0$. The difference in false positive rate between analysis and evaluations is $0.01$ in~\ham~and $0.02$ in the network of~\fb,~\wiki~and~\astro. We get exactly same results from our analytical model and simulation in the social network datasets of~\flic~and~\cat~(the largest network in our evaluations).

\section{Simulation and Results} \label{results}

In this section, we describe the design of our simulation and present the results of our experiments. We build our Sybil detection mechanism on top of Persea system and inherit hierarchical node ID distribution and certification, routing table organization and replication mechanisms of Persea~\cite{persea}. We incorporate our Sybil-filtering scheme with the lookup mechanism of Persea to achieve higher lookup success rate. 

We evaluate for a $31$-bit ID space. For different system parameters, we use the same values as used in Persea, such as: chunk-factor $c_f=0.65$, redundancy $R=7$, and Kad parameters $\alpha=5$, $\beta=7$, and bucket size $k=7$. In our evaluations, we assume that each node has one collaborative friend at each of its upper layers. 

We simulate for networks with different clustering coefficients and the experimental results show that the effectiveness of our Sybil detection scheme and the lookup success rate do not depend on the fast-mixing nature of a network. We represent the number of attack edges by $g$ and the number of benign peers by $n$. To validate our claim that iPersea gains high lookup success rate, even for a high value of $g/n$, we evaluate for this ratio, upto $1.5$.

\begin{table}[b]
\renewcommand{\arraystretch}{1.3}
\caption{Rate of false-positive for varying $g/n$ [Trusted collaborative friends]}
\centering
\begin{tabular}{|c|c|c|c|c|c|c|}
\hline
g/n &$0.10$ & $0.50$ & $0.80$ & $1.0$ & $1.25$ & $1.50$\\
\hline
\hline
\ham~(0.08)   & $0.046$ & $0.047$ & $0.049$ & $0.05$ & $0.09$ & $0.095$\\
\hline
\fb~(0.15)   & $0.063$ & $0.064$ & $0.066$ & $0.067$ & $0.069$ & $0.07$\\
\hline
\flic~(0.17)  & $0.037$ & $0.039$ & $0.04$ & $0.042$ & $0.048$ & $0.055$\\
\hline
\wiki~(0.21)	& $0.07$ & $0.078$ &$0.08$ & $0.09$ & $0.092$ & $0.093$ \\
\hline
\cat~(0.43) 	& $0.077$ & $0.079$ & $0.08$ & $0.082$ & $0.09$ & $0.12$ \\
\hline
\astro~(0.63)  & $0.06$ & $0.063$ & $0.064$ & $0.067$ & $0.09$ & $0.11$  \\
\hline
\end{tabular}
\label{tab:fp_trust}
\end{table}

\begin{table}[b]
\renewcommand{\arraystretch}{1.3}
\caption{Rate of false-positive for varying $g/n$ [Randomly selected collaborative friends]}
\centering
\begin{tabular}{|c|c|c|c|c|c|c|}
\hline
g/n &$0.10$ & $0.50$ & $0.80$ & $1.0$ & $1.25$ & $1.50$\\
\hline
\hline
\ham~(0.08)   & $0.046$ & $0.047$ & $0.08$ & $0.09$ & $0.14$ & $0.19$\\
\hline
\fb~(0.15)   & $0.063$ & $0.064$ & $0.067$ & $0.068$ & $0.071$ & $0.073$\\
\hline
\flic~(0.17)  & $0.037$ & $0.039$ & $0.04$ & $0.045$ & $0.052$ & $0.06$\\
\hline
\wiki~(0.21)	& $0.07$ & $0.078$ &$0.09$ & $0.121$ & $0.124$ & $0.126$ \\
\hline
\cat~(0.43) 	& $0.077$ & $0.08$ & $0.084$ & $0.087$ & $0.097$ & $0.13$ \\
\hline
\astro~(0.63)  & $0.06$ & $0.07$ & $0.079$ & $0.083$ & $0.108$ & $0.13$  \\
\hline
\end{tabular}
\label{tab:fp_rand}
\end{table}


\begin{table}[b]
\renewcommand{\arraystretch}{1.3}
\caption{Rate of false-negative for varying $g/n$ [Randomly selected collaborative friends]}
\centering
\begin{tabular}{|c|c|c|c|c|c|c|}
\hline
g/n &$0.10$ & $0.50$ & $0.80$ & $1.0$ & $1.25$ & $1.50$\\
\hline
\hline
\ham~(0.08)   & $0.0$ & $0.0$ & $0.031$ & $0.04$ & $0.05$ & $0.095$\\
\hline
\fb~(0.15)   & $0.0$ & $0.0$ & $0.001$ & $0.001$ & $0.002$ & $0.003$\\
\hline
\flic~(0.17)  & $0.0$ & $0.0$ & $0.0$ & $0.003$ & $0.004$ & $0.005$\\
\hline
\wiki~(0.21)	& $0.0$ & $0.0$ &$0.01$ & $0.031$ & $0.032$ & $0.033$ \\
\hline
\cat~(0.43) 	& $0.0$ & $0.001$ & $0.004$ & $0.005$ & $0.007$ & $0.01$ \\
\hline
\astro~(0.63)  & $0.0$ & $0.007$ & $0.015$ & $0.016$ & $0.018$ & $0.02$  \\
\hline
\end{tabular}
\label{tab:fn_rand}
\end{table}

\begin{figure*}[!t]
\centering
\subfigure[hamsterster (0.08)]{
\label{fig:look_ham}
\includegraphics[width=55mm,height=38mm]{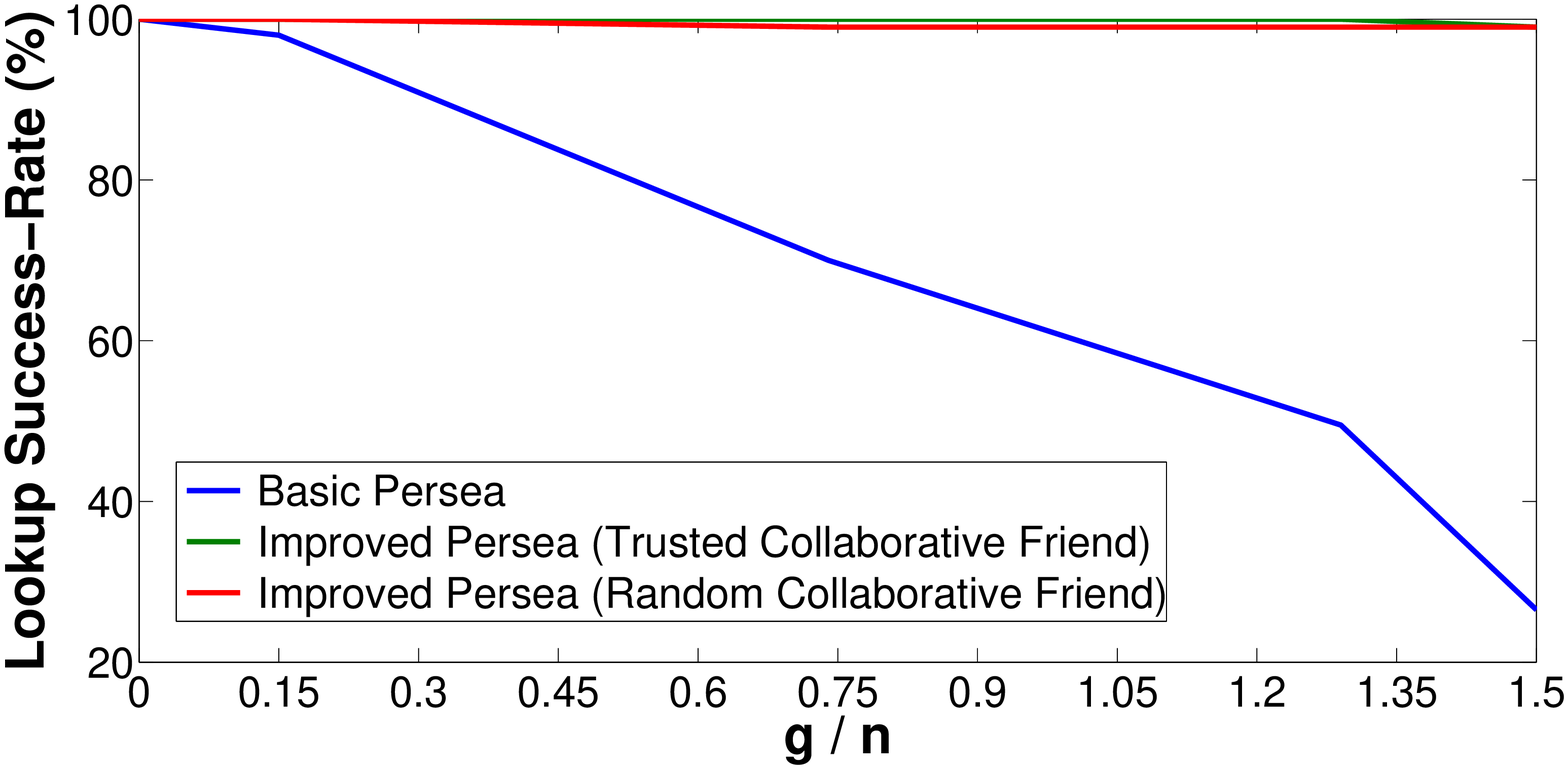}}\hspace{0.4cm}
\subfigure[facebook (0.15)]{
\label{fig:look_fb}
\includegraphics[width=55mm,height=38mm]{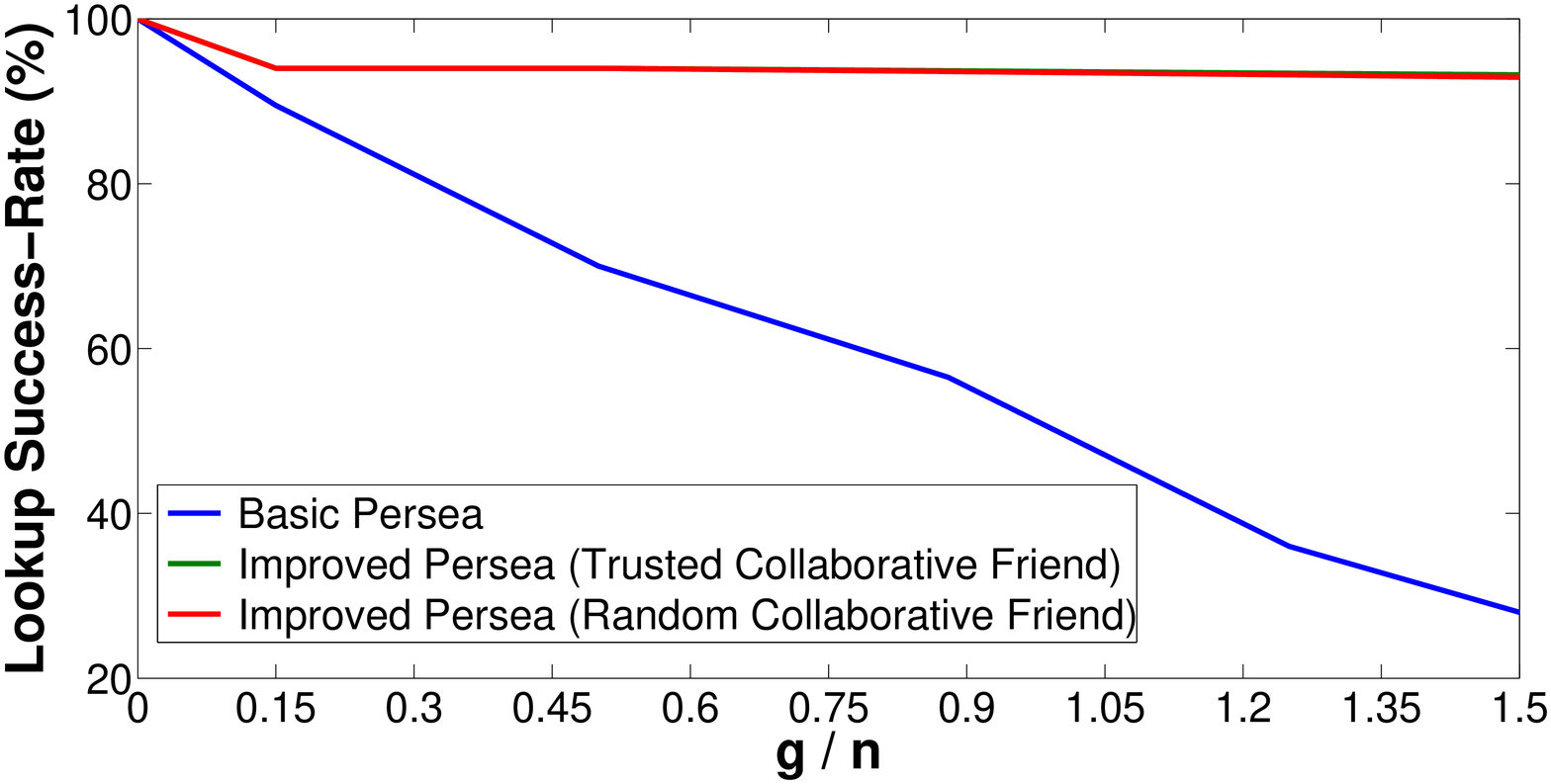}}
\subfigure[flickr (0.17)]{
\label{fig:look_flic}
\includegraphics[width=55mm,height=38mm]{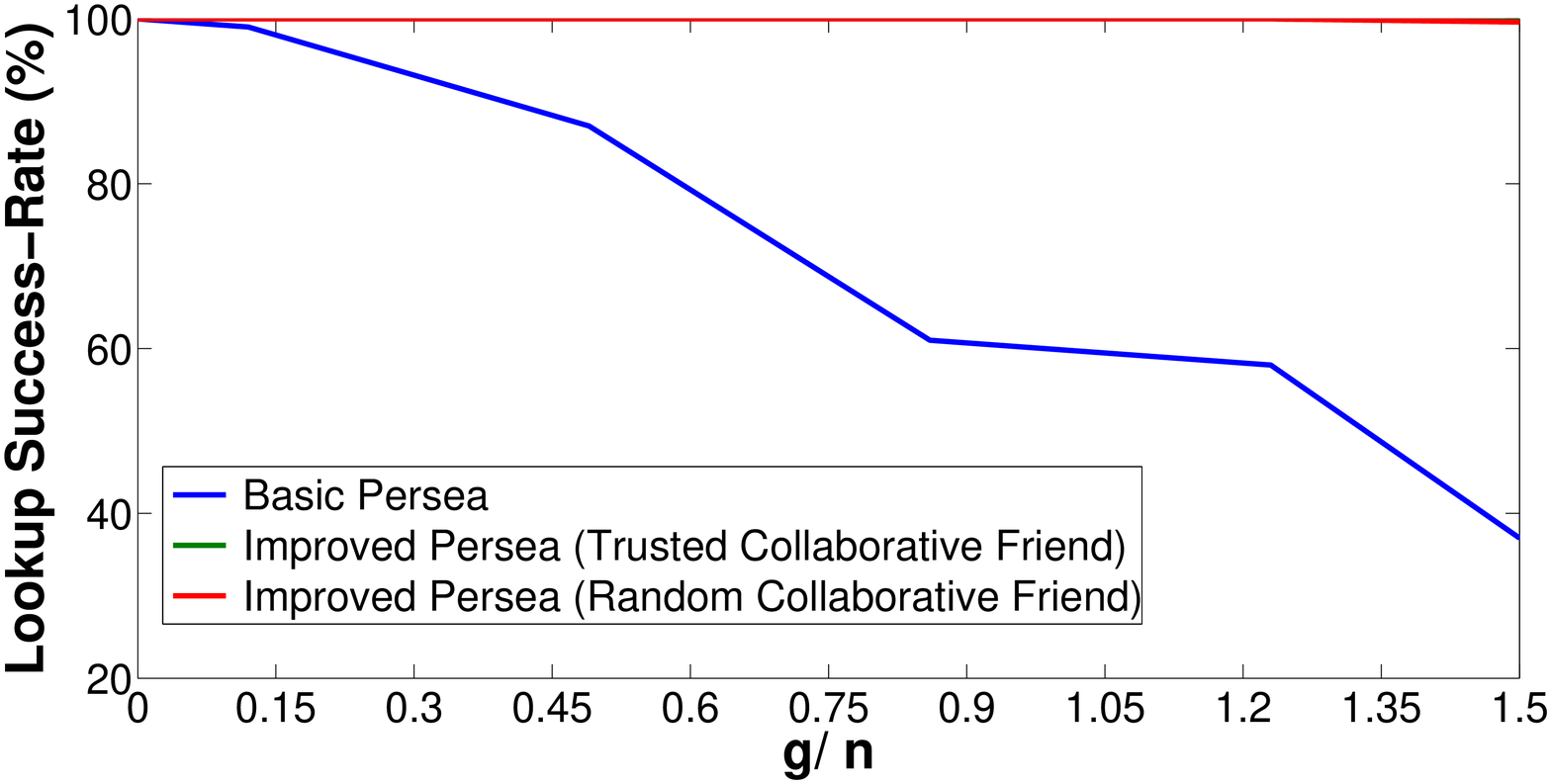}}\hspace{0.4cm}
\subfigure[wiki-Vote (0.21)]{
\label{fig:look_wiki}
\includegraphics[width=55mm,height=38mm]{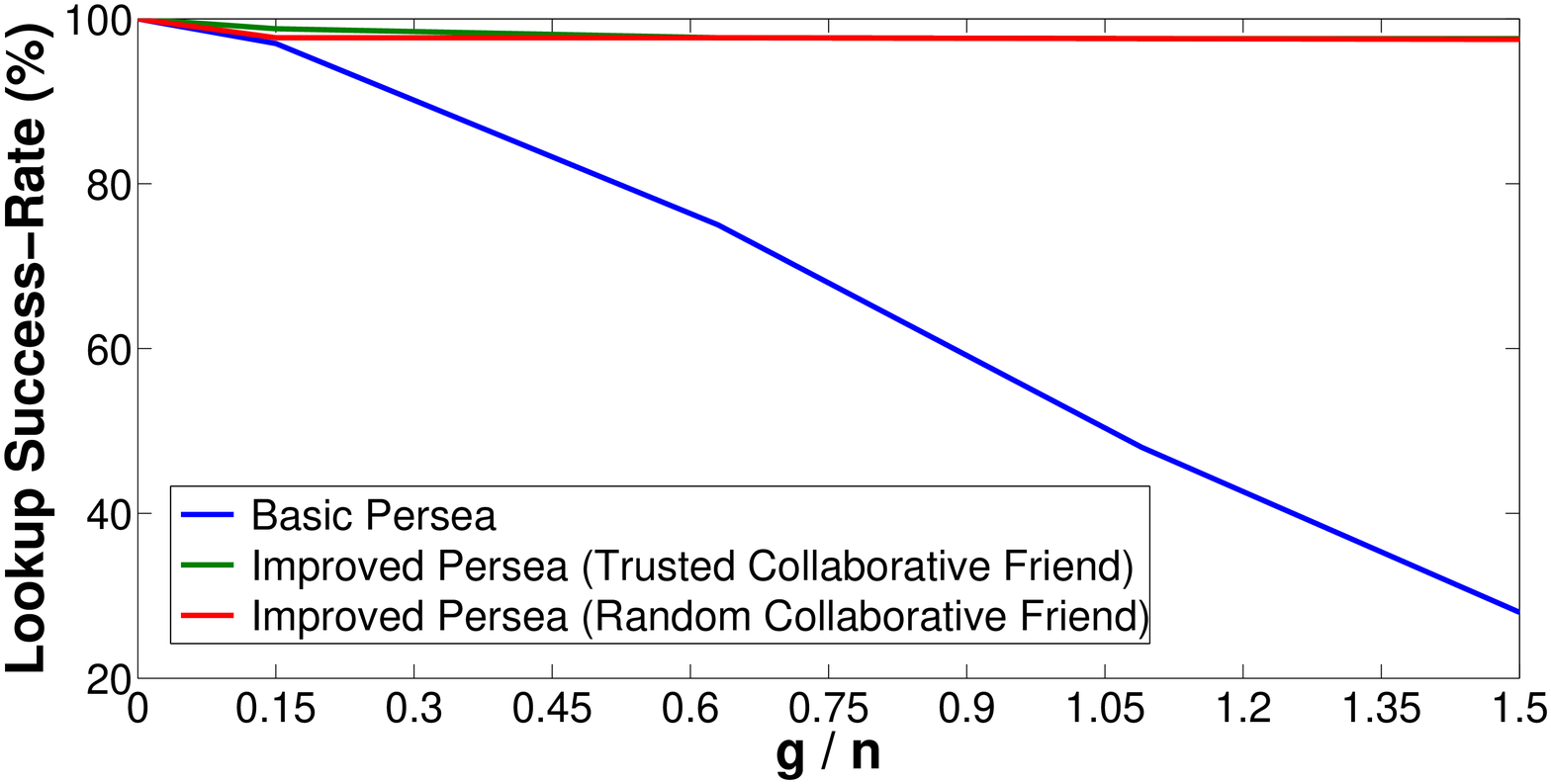}}
\subfigure[catster (0.43)]{
\label{fig:look_cat}
\includegraphics[width=55mm,height=38mm]{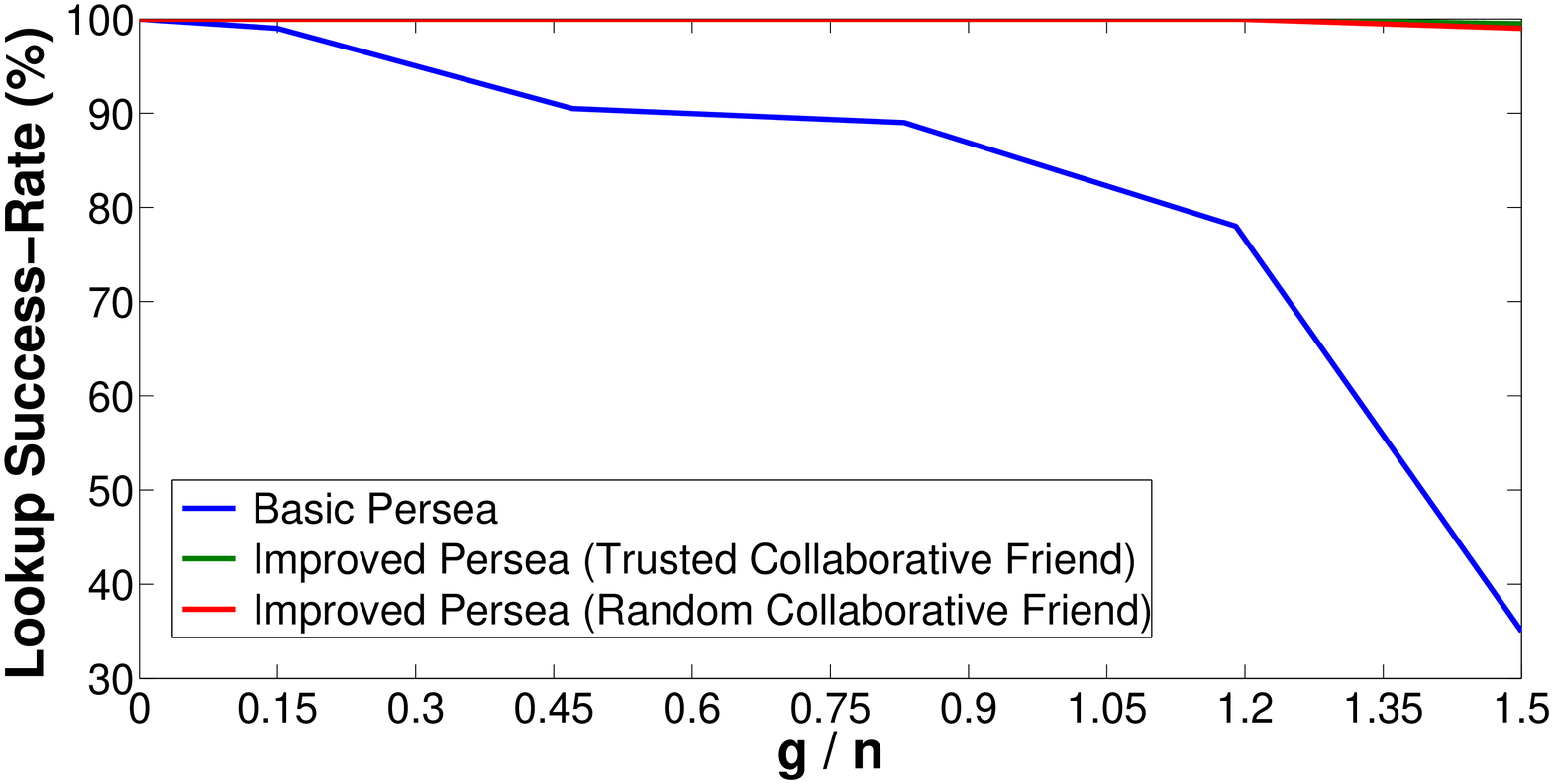}}
\subfigure[ca-AstroPh (0.63)]{
\label{fig:look_astro}
\includegraphics[width=55mm,height=38mm]{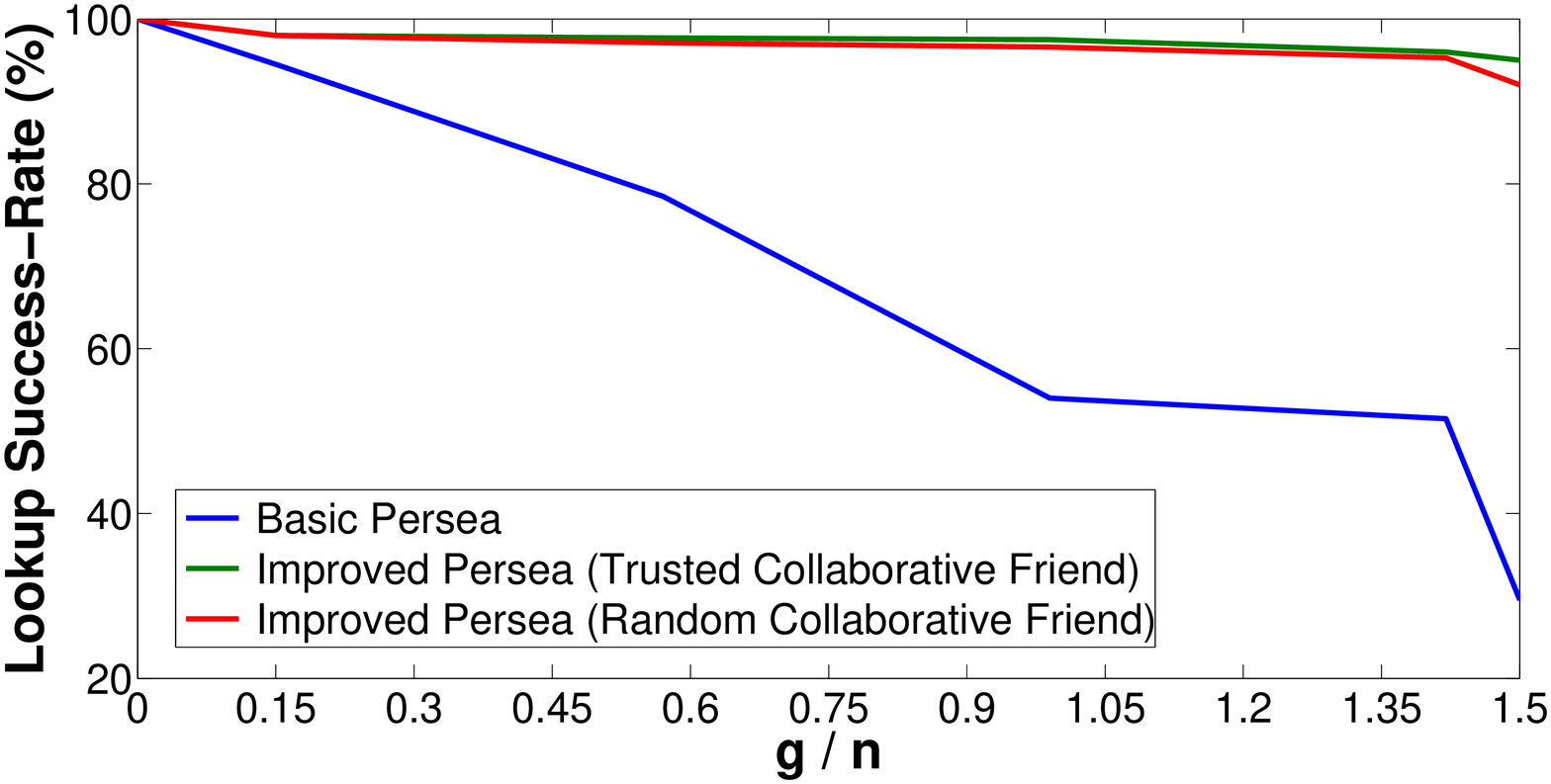}}\hspace{0.4cm}
\caption{Lookup success rates in networks with different clustering-coefficients}
\label{fig:success}
\end{figure*}

\subsection{Building the Network and Joining of Attackers}\label{build}

We follow the exact same approach, as in Persea~\cite{persea}, for building the network and joining of attackers. So, we build the bootstrap tree by emulating the process of nodes joining via existing connections in a social network graph. Although our system does not rely on the structure of the social graph for its security properties, we use real social network graphs to provide a realistic basis for the choices that nodes make in building the tree.

As we construct the bootstrap graph, the nodes in these datasets are considered to be honest and attacker nodes are further added to the network. We implement Persea system with a deployment that starts with seven randomly selected bootstrap nodes. We then use breadth-first-search over the social graph to add other nodes. After adding all of the honest nodes, we add Sybil nodes by creating attack edges to randomly selected honest peers. 

We evaluate our mechanism in simulations for one collaboration network : ca-AstroPh (\astro) and five social network datasets: facebook (\fb), flickr (\flic), catster (\cat), wiki-Vote (\wiki) and hamsterster (\ham). Here,~\wiki~uses directed edges to indicate ``who trusts whom'', which we believe is a good proxy for the notion that parent node would accept another node as a child in the bootstrap graph. In our evaluations,\ham,\fb,\flic~and~\cat~are social networks drawn from the users on Hamsterster.com, Facebook.com, Flickr.com and Catster.com Websites, respectively. 

In our evaluations,~\cat~is the largest network with $149700$ nodes and $5449275$ edges. While considering clustering coefficients,~\ham~and~\astro~are the networks with the smallest ($0.08$) and largest ($0.63$) clustering coefficient, respectively. Table~\ref{tab:topology} shows the sizes and clustering coefficients of the network datasets \footnote{\url{http://snap.stanford.edu/data}, \url{http://konect.uni-koblenz.de}, \\ \url{http://socialcomputing.asu.edu/pages/datasets}, }.

\begin{table}[b]
\renewcommand{\arraystretch}{1.3}
\caption{Average hop-count in inspection lookup [$g/n = 1.5$]}
\centering
\begin{tabular}{|c|c|c|}
\hline
Network & Trusted  & Random\\
         & collaborative friend  & collaborative friend\\
\hline
\hline
\ham~(0.08)   & $1.27$ & $1.35$ \\
\hline
\fb~(0.15)   & $1.29$ & $1.29$ \\
\hline
\flic~(0.17) & $1.41$ & $1.42$ \\
\hline
\wiki~(0.21)	& $1.24$  & $1.26$\\
\hline
\cat~(0.43)	& $1.69$ & $1.71$\\
\hline
\astro~(0.63)  & $1.10$ & $1.11$ \\
\hline
\end{tabular}
\label{tab:hpc}
\end{table}

\begin{table}[b]
\renewcommand{\arraystretch}{1.3}
\caption{Average hop-count in regular lookup [$g/n = 1.5$]}
\centering
\begin{tabular}{|c|c|c|c|}
\hline
Network &Persea & iPersea (Trusted  & iPersea (Random\\
  &   & collaborative friend)  & collaborative friend)\\
\hline
\hline
\ham~(0.08)   & $2.85$ & $2.80$ &$2.84$\\
\hline
\fb~(0.15)      & $4.79$ & $4.70$ & $4.70$\\
\hline
\flic~(0.17)     & $3.84$ & $3.44$ & $3.50$\\
\hline
\wiki~(0.21)	& $3.24$ & $3.18$ & $3.20$ \\
\hline
\cat~(0.43)	& $3.69$ & $3.59$ &$3.62$\\
\hline
\astro~(0.63)  & $3.93$ & $3.87$ & $3.88$\\
\hline
\end{tabular}
\label{tab:overhead}
\end{table}

\subsection{Rate of False Positive and False Negative} In this section, we show the results for the rate of false positive and false negative in our Sybil detection mechanism, when the collaborative friends are either trusted or randomly selected.

When trusted nodes are selected as collaborative friends, the results in Table~\ref{tab:fp_trust} show that for $g/n=1.0$, the rate of false positive is $0.05$ in \ham~(network smallest clustering coefficient) and $0.067$ in \astro~(network largest clustering coefficient). When the number of attack edges is equal to the number of honest peers, the maximum rate of false positive for any network is found $0.09$ in \wiki. 

For the random selection of collaborative friends, the rate of false positive slightly increases as compared to the trusted collaborative friends; for example, in~\cat, the largest network in our experiments, when $g/n=1.0$, the rates of false positive are $0.082$ and $0.087$ for trusted and randomly selected collaborative friends, respectively. Table~\ref{tab:fp_rand} shows the rates of false positive for randomly selected collaborative friends. 

The rate of false negative is zero for trusted collaborative friends. However, for randomly selected collaborative friends, the rates of false negative vary in the range between $0.001$ (\fb) and $0.04$ (\ham), when the ratio of attack edges to the number of honest peers is one (shown in Table~\ref{tab:fn_rand}). The rate of false negative gets much lower for decreasing $g/n$; for example, in~\flic, the rate of false negative is zero when $g/n=0.80$.

\subsection{Lookup Success Rate} We evaluate Persea with majority voting scheme and then compare the lookup success rate with iPersea for networks with different clustering coefficients, shown in Figure~\ref{fig:success}. The results show that iPersea performs much better than Persea. For increasing $g/n$, lookup success rate in Persea decreases sharply. We evaluate iPersea for $g/n$ upto $1.5$ and find consistent results for increasing $g/n$; in iPersea, the lookup success rate for any network in our simulation is no less than $92.9$\%. We get $100$\% lookup success rate in~\cat~and~\flic~for both trusted and randomly selected collaborative friends, even when the number of attack edges is equal to the number of honest peers in our system.

When $g/n = 1.0$ in~\ham~(network with lowest clustering coefficient), the lookup success rate in Persea is $59$\%, whereas the percentage of successful lookup in iPersea is $100$\% for trusted collaborative friends and $99$\% for randomly selected collaborative friends. In~\astro~(network with largest clustering coefficient), $54$\% lookups succeed in Persea when $g/n=1.0 $. For the same ratio, the lookup success rates in iPersea are $97.5$\% and $96.6$\% for trusted and randomly selected collaborative friends, respectively.

Our experimental results show that the lookup success rates for trusted and randomly selected collaborative friends remain same in~\cat,~\flic~and~\wiki, when the number of attack edges is equal to the number of honest peers. For the same ratio of attack edges to honest peers, while comparing these two approaches of selecting collaborative friends, the differences in the percentage of successful lookup in other networks are as follows: $0.04$\% in~\fb, $0.09$\% in~\astro~and $1$\% in~\ham. Hence, the lookup success rates for random selection of collaborative friends are very close to that in the ideal scenario, which assumes, the collaborative friends are trustworthy.

\subsection{Overhead} We evaluate to figure out the overhead of our system in terms of average hop-count per lookup. Our experimental results show the overhead for both inspection and regular lookups, when the collaborative friends are either trusted or randomly selected. We also compare our overheads for regular lookup with Persea.

{\em Inspection lookup:} Let us assume, the role of node C as the intermediate hop gets inspected. In ideal case, node C has the target of the lookup in its friend-list and in this scenario, the number of intermediate peer to reach the target is one. However, hop-count increases if the target node is not a direct friend of node C. Table~\ref{tab:hpc} shows the results for average hop-count per inspection lookup. 

The results illustrate that when the trusted nodes are selected as collaborative friends, average hop-count for different networks vary in the range between $1.10$~(\astro) and $1.69$~(\cat). For random selection of collaborative friends, the minimum and maximum average hop-counts per inspection lookup are $1.11$~(\astro) and $1.71$~(\cat), respectively. Hence, the differences between these two approaches of selecting collaborative friends are quite subtle, in terms of average hop-count per inspection lookup. 

{\em Regular lookup:} Our evaluations figure out the overheads associated with regular lookups, shown in Table~\ref{tab:overhead}, which also represent the overhead for a peer to get the status of a node from its parent. The increase in hop-count is found very small, when the collaborative friends are randomly selected instead of selecting the trusted peers only.

The results show that iPersea achieves some improvements over Persea while considering the average hop-count per regular lookup. In both Persea and iPersea, the maximum hop-count per regular lookup is found in~\fb, which is $4.79$ in Persea and $4.70$ in iPersea for both trusted and random collaborative friends. In~\cat, the largest network of our simulation, average hop-count per regular lookup is $3.69$ in Persea, which $3.59$ for trusted collaborative friends and $3.62$ for randomly selected collaborative friends in iPersea.

\section{Related Work} \label{related}

Sybil attacks can be leveraged to greatly undermine the operations of a variety of systems, including P2P systems (as examined in this paper), online social networks (OSNs), wireless sensor networks, online ratings systems, and more. Because of the power and generality of the Sybil attack, a large number of defenses have been proposed~\cite{survey_sybil}. Since we have already discussed about Persea, in this section we examine other major approaches proposed in literature that use social network in their Sybil defense mechanisms.

A number of works have proposed Sybil detection techniques or Sybil resistance based on random walks over a social network~\cite{whanau1,whanau2,xvine,sybillimit,sybilguard,sybilinfer,sybildefender}. The basic idea is that we can divide the social network into a Sybil region and an honest region connected via a small number of attack edges (a {\em small cut}). Random walks starting from the honest region have a low probability of ending in the Sybil region. This can be leveraged in a variety of ways, leading to detection mechanisms~\cite{sybilguard,sybillimit,sybilinfer,sybildefender}, and Sybil-resistant P2P designs~\cite{whanau1,whanau2,xvine}.

All of these mechanisms require the absence of small cuts within the honest region in the underlying social network (i.e., the honest region should be fast-mixing). Experimental results of Mohaisen et al., however, show that the mixing time of many real social networks is slower than the mixing time assumed by these works~\cite{mixing}. Mohaisen et al. also point out that some of these works have made questionable assumptions in their evaluations, which may have helped lead to the good results that have been published for these schemes~\cite{mixing}. Additionally, many real-world social networks fail to satisfy the other requirements of the systems, either because a significant fraction of nodes are sparsely connected or the users are organized in small tightly-knit communities, which are sparsely interconnected~\cite{comp_sybil}.

Lesniewski-Laas proposes a protocol~\cite{whanau1} in which a node constructs its routing table through independent random walks and recording the final node in each walk as the finger in its routing table. The protocol~\cite{whanau1} is extended in~\cite{whanau2} where the idea of layered identifiers is introduced to counter clustering attacks.

As mentioned before, \whanau relies on the assumptions of a fast-mixing social network and small number of attack edges, which may not hold in real social networks~\cite{mixing,viswanath_analysis,sybil_wild,id_theft}. Lesniewski-Laas et. al.'s own results on mixing in real social networks~\cite{whanau2} still show a noticable gap from the expected result for a fast-mixing network. Also, \whanau requires significant routing table state on the order of $O(\sqrt{n}\log{n})$, where $n$ is the number of objects stored in the DHT. Mittal et al. point out that the network overhead for maintaining this state can be substantial (e.g. 800 KBps per node)~\cite{xvine}.

X-Vine~\cite{xvine} works by communicating over social network edges. It builds a DHT on the top of a social network, where each node in the system selects a random numeric identifier. In the identifier space, each node maintains paths to its neighbors. In X-Vine, honest peers rate-limit the number of paths that are allowed to be built over their adjacent edges, which helps to limit the number of Sybil nodes that can join the system.  X-Vine, however, relies on the fast-mixing assumption and was only evaluated with a small number of attack edges (one for every ten honest nodes).

\section{Future Work}\label{future}

We would extend our attacker detection scheme to handle the {\em oscillation attack}. In such attacks, an attacker performs as both honest and malicious peer at regular interval. So, if an inspection lookup is performed during its honest-role, the attacker may get '+' status. To prevent such attacks, the parent node continues to perform inspection lookup for the child with '+' status at random interval and once a node gets '-' status, it is sealed permanently, since we assume that an honest node always performs legitimately. 

The above mechanism is also effective in getting the current status of a node, whose role is changed from honest to malicious because of being compromised by an attacker. In our future work, we would implement the above strategies to make our system robust against such attacks.

\section{Conclusion} \label{conclusion}

In this paper, we propose a Sybil detection mechanism, which accommodates a specially designed lookup mechanism to detect the Sybils and a filtering mechanism to opt-out those detected attackers during regular DHT lookup. We incorporate our mechanisms with Persea to develop the system: iPersea, which inherits the advantages of Persea that it gains over prior systems, and ensures higher lookup success rate even when the malicious targets respond with incorrect results. Our Sybil detection mechanism can be amended for incorporating with any DHT, where the children of a node are connected through their parents, required to implement the inspection lookup.

\balance

\section{Acknowledgement}
This material is based upon work supported by the National Science Foundation under Grant No. CNS-1117866 and CAREER Grant No. 0954133.

\bibliographystyle{IEEEtran}      
\bibliography{IEEEabrv,refs}
\balance

\end{document}